\documentstyle{article}
\title{\LARGE Hamiltonian formalism in a problem of 3-th waves hierarchy}
\author{A.~N.~Leznov\thanks{ Universidad Autonoma del Estado de Morelos, CCICAp,Cuernavaca, Mexico}} \date{}

\newcommand{\rig}[2]{\stackrel{#2\rightarrow}{#1}}

\begin{document}
\maketitle

\maketitle

\begin{abstract}

By the method of discrete transformation equations of 3-th wave hierarchy are constructed.  
We present in explicit form two Poisson structures, which allow to construct Hamiltonian operator consequent application of which leads to all equations of this hierarchy.
For calculations it will be necessary results of previous paper \cite{1}, which for 
convenience of the reader we present in corresponding place of the text. The obtained formulae are checked by independent calculations.

\end{abstract}

\section{Introduction}

 All system equations of 3-th wave hierarchy are invariant with respect to two mutually commutative discrete transformation of this problem \cite{1},\cite{I},\cite{tok},\cite{nlm}\cite{1}. In this introduction we present the solution of the same problem in the case $A_1$ algebra follow to the paper \cite{DL}.  

We repeat here briefly the most important punks of general construction from \cite{DL}.

The discrete invertible substitution (mapping) defined as
\begin{equation}
\tilde u=T(u,u',...,u^{r})\equiv T(u)\label{1}
\end{equation}
$u$ is $s$ dimensional vector function; $u^r$ its derivatives of corresponding order with 
respect to "space" coordinates.

The property of invertibility means that (\ref{1}) can be resolved and "old" function $u$
may expressed in terms of new one $\tilde u$ and its derivatives.

Freshet derivative $T'(u)$ of (\ref{1}) is $s\times s$ matrix operator defined as
\begin{equation}
T'(u)=T_{u}+T_{u'}D+T_{u''}D^2+...\label{FR}
\end{equation}
where $D^m$ is operator of m-times differentiation with respect to space coordinates.

Let us consider equation
\begin{equation}
F_n(T(u))=T'(u)F_n(u)\label{ME}
\end{equation}
where $F_n(u)$ is s-component unknown vector function, each component of which depend on $u$
and its derivatives not more than $n$ order. It is not difficult to understand that evolution type equation
$$
u_t=F_n(u)
$$
is invariant with respect substitution (\ref{1}). 

Two other equations and its solutions are important in what follows
\begin{equation}
T'(u)J(u)(T'(u))^T=J(T(u)),\quad T'(u)H(u)(T'(u))^{-1}=H(T(u))\label{H}
\end{equation}
where $(T'(u))^T=T_{u}^T-DT_{u'}^T+D^2T^T_{u''}+...$ and $J(u),H(u)$ are unknown $s\times s$
matrix operators, the matrix elements of which are polynomial of some finite order with respect
to operator of differentiation (of its positive and negative degrees).

$J^T(u)=-J(u)$ may be connected with the Poisson structure and equation (\ref{H}) means its invariance with respect to discrete transformation $T$.

The second equation (\ref{H}) determine operator $H(u)$, which after application to arbitrary solution of (\ref{ME}) $F(u)$ leads to new solution of the same system
$$
\tilde F(u)=H(u)F(u)
$$
And thus we obtain reccurent procedure to construct solutions of (\ref{ME}) from few simple ones.
 
If it is possible to find two different $J_1,J_2$ (Hamiltonian operators,Poisson structures) then
\begin{equation}
H(u)=J_2J_1^{-1}\label{HH}
\end{equation}
satisfy second equation (\ref{H}).

In \cite{DL} are presented arguments that Hamiltonian operator it is the sense find in a form
\begin{equation}
J(u)=F_n(u)D^{-1}F_n(u)^T+\sum_{i} A_iD^i\label{J}
\end{equation}
where $F_n$ some solution of (\ref{ME}) and $A_i$ some $s\times s$ matrices constructed from
$u$ and its derivatives.

The direct generalization of (\ref{JJ}) which will be used below is the following
\begin{equation}
J(u)=\sum F_i(u)D^{-1}F_i(u)^T+\sum_{i} A_iD^i\label{JJ}
\end{equation}
where first term in (\ref{J}) is changed on sum of some number of different solutions of (\ref{ME}). 

\section{Necessary facts from \cite{1}}

In \cite{1} was constructed equations of 3-th waves hierarchy of zero, first and second order for six unknown functions $f^{\pm}_{1.0},f^{\pm}_{0.1},f^{\pm}_{1.1}$. The form
of these equations will be essentially used in what follows.

\subsection{Equations of the zero order}

\begin{equation}
\dot {f^{\pm}_{1.0}}=\pm bf^{\pm}_{1.0},\quad
\dot {f^{\pm}_{0.1}}=\pm cf^{\pm}_{1.0},\quad \dot {f^{\pm}_{1.1}}=\pm af^{\pm}_{1.1}\label{ABC}
\end{equation}
where $(b,c)$ arbitrary numerical parameters $a=b+c$.

\subsection{Equations of the first order}

$$
\dot {f^+_{1.1}}=\theta_{1.1}(f^+_{1.1})'+\sigma_{1.1}f^+_{1.0}f^+_{0.1},\quad
\dot {f^+_{1.0}}=\nu_{1.0}(f^+_{1.0})'+\sigma_{1.0}f^-_{0.1}f^+_{1.1}
$$
\begin{equation}
\dot {f^+_{0.1}}=\nu_{0.1}(f^+_{0.1})'+\sigma_{0.1}f^-_{1.0}f^+_{1.1},
\quad \dot {f^-_{0.1}}=\nu_{0.1}(f^-_{0.1})'+\sigma_{0.1}f^+_{1.0}f^-_{1.1}\label{FD}
\end{equation}
$$
\dot {f^-_{1.0}}=\nu_{1.0}(f^-_{1.0})'+\sigma_{1.0}f^+_{0.1}(f^-_{1.1},\quad
\dot {f^-_{1.1}}=\theta_{1.1}(f^-_{1.1})'+\sigma_{1.1}f^-_{1.0}f^-_{0.1}.
$$
where $\nu_{i,j},\sigma_{ij},\theta_{11}$ are numerical parameters connected by condition
\begin{equation}
\nu_{01}-\nu_{10}=2\sigma_{10},\quad \nu_{01}+\nu_{10}=2\theta_{11},\quad -2\sigma_{11}=\sigma_{10}=\sigma_{01}\label{EX}
\end{equation}
Thus solution is defined by two independent parameters $\nu_{01},\nu_{10}$ as in the case of zero degree solution of the previous subsection. 

\subsection{Equations of the second order}
  
$$
\dot {f^+_{1.1}}=\nu_{1.1}(f^+_{1.1})''+\gamma_{1.1}f^+_{1.0}(f^+_{0.1})'+\delta_{1.1}f^+_{0.1}
(f^+_{1.0})'+f^+_{1.1}R_{11},
$$
where $a\equiv (b+c),R_{ij}\equiv 2a_{ij}f^+_{1.1}f^-_{1.1}+b_{ij}f^+_{1.0}f^-_{1.0}+c_{ij}f^+_{0.1}f^-_{0.1}$.
$$
\dot {f^+_{1.0}}=\nu_{1.0}(f^+_{1.0})''+\gamma_{1.0}f^-_{0.1}(f^+_{1.1})'+\delta_{1.0} f^+_{1.1}(f^-_{0.1})'+f^+_{1.0}R_{10}
$$
$$
\dot {f^+_{0.1}}=\nu_{0.1}(f^+_{0.1})''+\gamma_{0.1}f^-_{1.0}(f^+_{1.1})'+\delta_{0.1}f^+_{1.1}(f^-_{1.0})'+f^+_{0.1}R_{01},
$$
\begin{equation}
{}\label{21}
\end{equation}
$$
-\dot {f^-_{0.1}}=\nu_{0.1}(f^-_{0.1})''+\gamma_{0.1}f^+_{1.0}(f^-_{1.1})'+\delta_{0.1}f^-_{1.1}(f^+_{1.0})'+f^-_{0.1}R_{01},
$$
$$
-\dot {f^-_{1.0}}=\nu_{1.0}(f^-_{1.0})''+\gamma_{1.0}f^+_{0.1}(f^-_{1.1})'+\delta_{1.0}f^-_{1.1}(f^+_{0.1})'+f^-_{1.0}R_{10},
$$
$$
-\dot {f^-_{1.1}}=\nu_{1.1}(f^-_{1.1})''+\gamma_{1.1}f^-_{1.0}(f^-_{0.1})'+\delta_{1.1}f^-_{0.1}(f^-_{1.0})')+f^-_{1.1}R_{11}.
$$
All numerical parameters in (\ref{21}) may be expressed in terms of only two ones and are connected by relations
$$
\nu_{11}=a_{11},\quad \gamma_{1.1}+\delta_{1.1}=(b_{11}-c_{11}),\quad \nu_{10}=2b_{10}
$$
$$
\gamma_{1.0}+\delta_{1.0}=2(c_{10}-b_{10}),\quad \nu_{01}=2c_{01},\quad \gamma_{0.1}+\delta_{0.1}=2(c_{01}-b_{01})
$$
\begin{equation}
H_M=\pmatrix{ a_{11}=-2c_{10} & b_{11}=a_{10} & c_{11}=-3c_{10}-b_{10} \cr
          a_{10}=c_{10}+b_{10} & b_{10}=b_{10} & c_{10}=c_{10} \cr
         a_{01}=-3c_{10}-b_{10} & b_{01}=c_{10} & c_{01}=-b_{10}-4c_{10} \cr}\label{HM}
\end{equation} 
$$
\delta_{10}=4c_{10},\quad \gamma_{10}=-2(c_{10}+b_{10}),\quad 2\gamma_{11}=\delta_{10}-\gamma_{10},
$$
$$
2\delta_{11}=-\gamma_{10},\quad \delta_{01}=-\delta_{10},\quad \gamma_{01}=\gamma_{10}-\delta_{10}
$$

\subsection{Hamiltonian form of equations}

As it was shown in (\ref{1}) equations of the previous subsections may be considered as Hamiltonian ones with following non zero Poisson breakets
\begin{equation}
\{f^+_{1.1},f^-_{1.1}\}={1\over 2},\quad \{f^+_{1.0},f^-_{1.0}\}=1,\quad \{f^+_{0.1},f^-_{0.1}\}=1\label{P}
\end{equation}
This fact leads to existence of the first Poisson structure, inverse to which is the following 
\begin{equation}
J_1^{-1}=\pmatrix{0 & 0 & 0 & 0 & 0 & - 2 \cr   
             0 & 0 & 0 & 0 & -1 & 0 \cr
             0 & 0 & 0 & -1 & 0 & 0 \cr
             0 & 0 & 1 & 0 & 0 & 0 \cr
             0 & 1 & 0 & 0 & 0 & 0 \cr
             2 & 0 & 0 & 0 & 0 & 0 \cr}\label{J1}
\end{equation}

\section{Hamiltonian operator of 3-th waves problem}

Let us seek in connection the proposition (\ref{JJ}) of introduction the second Poisson structure in a form  $J_2=J^n_2+J^p_2$, where $J^n_2$ contain terms with $D^{-1}$ and $J^p_2$ terms with non negative degree of $D$
$$
J^n_2=-F^1_0D^{-1}(F^1_0)^T-F^2_0D^{-1}(F^2_0)^T
$$
where $F^{1,2}_0$ are two different solutions of equations of the zero order (first subsection
of the previous section).
\begin{equation}
J^p_2=\pmatrix{0 & 0 & 0 & -{1\over 4}f^+_{1.0} & {1\over 4}f^+_{0,1} & {1\over 4}D \cr   
                       0 & 0  & f^+_{1.1} & 0 & D & -{1\over 4}f^-_{0.1} \cr
                       0 & -f^+_{1.1} & 0 & D & 0 & {1\over 4}f^-_{1.0} \cr
                       {1\over 4}f^+_{1.0} & 0 & D & 0 & -f^-_{1.1} & 0 \cr
                       -{1\over 4}f^+_{0.1} & D & 0 & f^-_{1.1} & 0 & 0 \cr
                   {1\over 4}D & {1\over 4}f^-_{0.1}& -{1\over 4}f^-_{1.0} & 0 & 0 & 0 \cr}\label{J2}
\end{equation}
In the last expression we present finally result. Really it is necessary to write anti symmetrical matrix with arbitrary coefficients, which will be found after calculations described below. 

Now let us consider how reccurent operator $H$ (\ref{HH}) acts on some solution of (\ref{ME}) $F$. At first
$F$ it is necessary multiply on $J^{-1}_1$ with the result
$$
J_1^{-1}F=\pmatrix{ -2F^-_{1.1} \cr
                     -F^-_{1.0} \cr
                     -F^-_{0.1} \cr
                     F^+_{0.1} \cr
                     F^+_{1.0} \cr
                    2F^+_{1.1} \cr}
$$
and this column vector multiply on $J_2=J^n_2+J^p_2$ from (\ref{J2}). In two terms of $J^n_2$
it is necessary multiply vector line $(F^i_0)^T$ on the last vector column with scalar result
$$
(F^i_0)^TJ_1^{-1}F=-2a^i(f^+_{1.1}F^-_{1.1}+f^-_{1.1}F^+_{1.1})-b^i(f^+_{1.0}F^-_{1.0}+f^-_{1.0}F^+_{1.0})-c^i(f^+_{0.1}F^-_{0.1}+f^+_{0.1}F^-_{0.1})
$$ 
Thus input of two first terms of $J^n_2$ into "new solution" will be
$$
\tilde {F^+_{1.1}}=-f^+_{1.1}D^{-1}(2a^2(f^+_{1.1}F^-_{1.1}+f^-_{1.1}F^+_{1.1})+(ab)(f^+_{1.0}F^-_{1.0}+f^-_{1.0}F^+_{1.0})+(ac)(f^+_{0.1}F^-_{0.1}+f^+_{0.1}F^-_{0.1}))
$$
\begin{equation}
\tilde {F^+_{1.0}}=-f^+_{1.0}D^{-1}(2(ba)(f^+_{1.1}F^-_{1.1}+f^-_{1.1}F^+_{1.1})+b^2(f^+_{1.0}F^-_{1.0}+f^-_{1.0}F^+_{1.0})+(bc)(f^+_{0.1}F^-_{0.1}+f^+_{0.1}F^-_{0.1}))\label{AK}
\end{equation}
$$
\tilde {F^+_{0.1}}=-f^+_{0.1}D^{-1}(2(ca)(f^+_{1.1}F^-_{1.1}+f^-_{1.1}F^+_{1.1})+(cb)(f^+_{1.0}F^-_{1.0}+f^-_{1.0}F^+_{1.0})+c^2(f^+_{0.1}F^-_{0.1}+f^+_{0.1}F^-_{0.1}))
$$
and the same expressions with opposite sign for components with negative upper indexes.
$a^2=\sum a^ia^i, (ab)=\sum a^ib^i$ and so on.

The result of multiplication $J^p_2J_1^{-1}F$ is determined by usual rules of multiplication matrix on vector
\begin{equation}
J^p_2J_1^{-1}F=\pmatrix{ {1\over 2}(F^+_{1.1})'+{1\over 4}(f^+_{0.1}F^+_{1.0}-f^+_{1.0}F^+_{0.1})\cr
(F^+_{1.0})'-f^+_{1.1}F^-_{0.1}-{1\over 2}f^-_{0.1}F^+_{1.1})\cr
(F^+_{0.1})'+f^+_{1.1}F^-_{1.0}+{1\over 2}f^-_{1.0}F^+_{1.1})\cr
-(F^-_{0.1})'-f^-_{1.1}F^+_{1.0}-{1\over 2}f^+_{1.0}F^-_{1.1})\cr
-(F^+_{1.0})'+f^-_{1.1}F^+_{0.1}+{1\over 2}f^+_{0.1}F^-_{1.1})\cr
-{1\over 2}(F^-_{1.1})'-{1\over 4}(f^-_{0.1}F^-_{1.0}-f^-_{1.0}F^-_{0.1})\cr}\label{BL}
\end{equation}

Now let us take for $F$ right hand side zero degree equations (subsection 1 from previous
section) $F^{\pm}_{1.1}=\pm(\nu_{1.0}+\nu_{0.1})f^{\pm}_{1.1},F^{\pm}_{1.0}=\pm\nu_{1.0}f^{\pm}_{1.0},
F^{\pm}_{0.1}=\pm\nu_{0.1}f^{\pm}_{0.1}$, ($b=\nu_{1.0},c=\nu_{0.1}$).
In this case input from $J^n_2$ terms (\ref{AK}) equal to zero and input from $J^p_2$ terms
exactly coincides with right hand side of equations of the first order (subsection 2 from previous section). Really calculations must be done in a back direction: 
in definition of $J^p_2$ (\ref{J2}) it is necessary to use arbitrary skew symmetrical matrix
and after comparison result of calculations above with first order equations obtain finally form $J^p_2$ (\ref{J2}).

Now we repeat the same trick with equations of the first degree. In this case
$$
J_1^{-1}F=
\pmatrix{ -(\nu_{1.0}+\nu_{0.1})(f^-_{1.1})'-{\nu_{1.0}-\nu_{0.1}\over 2}f^-_{1.0}f^-_{0.1} \cr
                     -\nu_{1.0}(f^-_{1.0})'+{\nu_{1.0}-\nu_{0.1}\over 2}f^-_{1.1}f^+_{0.1} \cr
-\nu_{0.1}(f^-_{0.1})'+{\nu_{1.0}-\nu_{0.1}\over 2}f^-_{1.1}f^+_{1.0} \cr
\nu_{0.1}(f^+_{0.1})'-{\nu_{1.0}-\nu_{0.1}\over 2}f^+_{1.1}f^-_{1.0} \cr
\nu_{1.0}(f^+_{1.0})'-{\nu_{1.0}-\nu_{0.1}\over 2}f^+_{1.1}f^-_{0.1} \cr
(\nu_{1.0}+\nu_{0.1})(f^+_{1.1})'+{\nu_{1.0}-\nu_{0.1}\over 2}f^+_{1.0}f^+_{0.1} \cr}
$$
We present result of action $J^p_2J_1^{-1}F$ below
$$
{\nu_{1.0}+\nu_{0.1}\over 4}(f^+_{1.1})''+{3\nu_{1.0}-\nu_{0.1}\over 8}(f^+_{1.0})'f^+_{0.1}+
{\nu_{1.0}-3\nu_{0.1}\over 8}f^+_{1.0})(f^+_{0.1})'+
$$
\begin{equation}
{\nu_{1.0}-\nu_{0.1}\over 8}f^+_{1.1}(f^-_{1.0}f^+_{1.0}-f^-_{0.1}f^+_{0.1})\label{FIN}
\end{equation}
The terms in the first line exactly coincide with terms with derivatives in equation of the second order for $f^+_{1.1}$ component. Indeed $(\nu_{1.0}+\nu_{0.1})=2b_{10}+2c_{01}=-8c_{10}=4a_{11}=4\nu_{11}$, $b_{10}={\nu_{10}\over 2},c_{10}=-{\nu_{1.0}+\nu_{0.1}\over 8}, \delta_{11}=b_{10}+c_{10}={3\nu_{1.0}-\nu_{0.1}\over 8}$ and so on. The same situation takes
place with respect all other components: terms with derivatives coincide with calculated from $J^p_2J_1^{-1}F$.  

Terms without derivatives arise from terms of the second line of (\ref{FIN}) and from (\ref{AK}). In the last equations after substitution $F$ from equation of the first order
(and in all other cases) under the sign $D^{-1}$ arises sign $D$ which lead to unity and
(\ref{AK}) plus terms of second line of (\ref{FIN}) look as
$$
f^+_{1.1}(2\theta_{11}a^2f^+_{1.1}f^-_{1.1}+[(ab)\nu_{10}+{\nu_{1.0}-\nu_{0.1}\over 8}]f^+_{1.0}f^-_{1.0}+[(ac)\nu_{01}-{\nu_{1.0}-\nu_{0.1}\over 8}]f^+_{0.1}f^-_{0.1}
$$
\begin{equation}
f^+_{1.0}([2(ba)\theta_{11}+{\nu_{1.0}-\nu_{0.1}\over 2]}f^+_{1.1}f^-_{1.1}+b^2\nu_{10}f^+_{1.0}f^-_{1.0}+[(bc)\nu_{01}-{\nu_{1.0}-\nu_{0.1}\over 8}]f^+_{0.1}f^-_{0.1})\label{AKK}
\end{equation}
$$
f^+_{0.1}([2(ca)\theta_{11}-{\nu_{1.0}-\nu_{0.1}\over 2}]f^+_{1.1}f^-_{1.1}+[(cb)\nu_{10}+{\nu_{1.0}-\nu_{0.1}\over 8}]f^+_{1.0}f^-_{1.0}+c^2\nu_{01}f^+_{0.1}f^-_{0.1})
$$
All terms above with scalar products arise from (\ref{AK}). All others from the "second line"
of (\ref{FIN}). Now it is necessary compleat these expressions with terms $f^+_{ij}R_{ij}$ of the right hand side of equation of the second order (subsection 3 of the previous section). 
This comparison leads to the following conclusion:
$$
a^2=b^2=c^2={1\over 2},\quad (ab)=(ac)=-(bc)={1\over 4}
$$
And now recurrent operator $H$ is defined uniquely. From this result it is clear that all attempts to construct $H$ using only one solution in the anzats for $J_2$ lead to contradiction 
and it was the main difficult to the author.

The following observation take place. Let us consider three elements of Cartan subalgebra
$R_3={h_1+h_2\over 2},R_2=h_2,R_1=h_1$ and the positive root system of $A_2$ algebra system $X^+_3=X^+_{12},X^+_2,X^+_1$. Let us define $3\times 3$ "Cartan matrix" $K^W$ by the condition
$$
[R_i,X^+_j]=K^W_{ji}X^+_j
$$
Then 
$$
\pmatrix{ a^2 & (ab) & (ac) \cr
                       (ba) & b^2 & (bc) \cr
                       (ca) & (cb) & c^2 \cr}={1\over 4}K^W
$$

\section{Hamiltonian formalism II}

In calculations of the previous section results of \cite{1} were used in the whole measure.
But the corresponding calculations were not simple and not strait forward. Now having the explicit expression for $J_2$ we are able to check that equation it defined (\ref{H}) is satisfied. Let us rewrite it notations of the previous section
$$
-T'(f)F^1_0D^{-1}(F^1_0)^T(T'(f))^T- T'(f)F^2_0D^{-1}(F^2_0)^T(T'(f))^T+
$$
\begin{equation}
T'(f)J^p_2(T'(f))^T=(J^n_2+J^p_2)(T(f))\label{HII}
\end{equation}
( we think that the same sign $T$ for substitution and transposition will not lead to mixing). 
We will do all calculations below with respect to $T_3$ transformation (explicit formulae for
it and $F'_3(f)$ reader can find in Appendix). We remind the rule of multiplication of quadratical in derivatives operator on scalar function $R$ 
\begin{equation}
(A+DB+DC^2)R=AR+BRA'+CRT''+(BRA+2CRT')D+CDR^2\label{R}
\end{equation}

Matrix elements of 3 first lines of $F'_3(f)$ do not contain operator of differentiation.
Its fourth and fifth lines linear in $D$ and its sixth one quadratical in $D$. By definition
(\ref{ME}) all terms without operator $D$ lead to $\rig{F^i_0}{T_3}$. All others can be simple
calculated using (\ref{R}) with the result 
\begin{equation}
T_3'(f)F^i_0=\rig{F^i_0}{T_3}+\pmatrix{0_3 \cr
                                   \Delta_if^-_{0.1}D \cr
                                  -\Delta_if^-_{1.0}D \cr
-{\Delta_i\over 2}f^-_{0.1}f^-_{1.0}D-a^if^-_{1.1}D^2 \cr}\equiv \rig{F^i_0}{T_3}+\rig{\Gamma^i}{T_3}\label{3}
\end{equation}
where $\Delta_i=c_i-b_i, a_i=c_i+b_i$, $0_3$- three dimensional zero vector
After substitution (\ref{3}) into (\ref{HII}) and cancelation equivalent terms in both sides
we come to the following equality have to be checked
$$
-\sum_{i=1}^{i=2}(\rig{F^i_0}{T_3}D^{-1}\rig{\Gamma^i}{T_3}^T+\rig{\Gamma^i}{T_3}D^{-1}\rig{F^i_0}{T_3}^T+\rig{\Gamma^i}{T_3}D^{-1}\rig{\Gamma^i}^T)+
$$
\begin{equation}
T'_3(f)J^p_2(T'_3(f))^T=J^p_2(T_3(f))\label{HIII}
\end{equation}
The matrix of the first sum in the first line has different from zero only elements of its last 
three columns. The second sum -  only elements of its three last line. And the last sum only elements of $3\times 3$ matrix in its left down corner.

At first let us calculate the first sum. Result is the following (we present below only two first lines of this matrix operator)
$$
-\sum_{i=1}^{i=2}(\rig{F^i_0}{T_3}D^{-1}\rig{\Gamma^i}{T_3}^T=
$$
\begin{equation}
\pmatrix{ 0_3 & 0 & 0 & {1\over 2}D+{1\over 2}{(f^-_{1.1})'\over f^-_{1.1}}\cr
0_3 & {3\over 4}{(f^-_{0.1})^2\over f^-_{1.1}} & -{3\over 4}{f^-_{0.1}f^-_{1.0}\over f^-_{1.1}} & -{3\over 8}{(f^-_{0.1})^2f^-_{1.0}\over f^-_{1.1}} -{1\over 4}f^-_{0.1}D -{1\over 4}{f^-_{0.1}(f^-_{1.1})'\over f^-_{1.1}} \cr}\label{FII}
\end{equation}
(where $0_3$ is three dimensional row vector)
$$
(T'_3(f)J^p_2(T'_3(f))^T)_{1l}=\sum_{m,k} (T'_3(f))_{1m}(J^p_2)_{mk}((T'_3(f))^T)_{kl}=
\sum_{k=1}^{3} (T'_3(f))_{16}(J^p_2)_{6k}(T'_3(f)^t)_{lk}=
$$
(because only one elements of the first line ($(T'_3(f))_{16}$) of Frechet derivative matrix and three elements of sixth line matrix $J^p_2$ are different from zero). 
$$
-{1\over (f^-_{1.1})^2}({1\over 4}D(T'_3(f))^t_{l1}+{1\over 4}f^-_{0.1}(T'_3(f))^t_{l2}-{1\over 4}f^-_{1.0}(T'_3(f))^t_{l3})
$$
where $(kD)^t\equiv -Dk^t$.
or the first line of the matrix $T'_3(f)J^p_2(T'_3(f))^T$ looks as
\begin{equation}
\pmatrix{ 0_3 & -{1\over 4}\rig{f^-_{1.0}}{T_3} & {1\over 4}\rig{f^-_{0.1}}{T_3} & -{1\over 4}D-{1\over 2}{(f^-_{1.1})'\over f^-_{1.1}}\cr}\label{f1}
\end{equation}

Now let us calculate second line
$$
(T'_3(f)J^p_2(T'_3(f))^T)_{2l}=\sum_{m,k} (T'_3(f))_{2m}(J^p_2)_{mk}((T'_3(f))^T)_{kl}=
$$
$$
\sum_{k=1}^{3} (T'_3(f))_{26}(J^p_2)_{6k}(T'_3(f)^t)_{lk}+\sum_{k=1}^{6} (T'_3(f))_{24}(J^p_2)_{4k}(T'_3(f)^t)_{lk}=
$$
(because only two elements $(T'_3(f))_{24},(T'_3(f))_{26}$ of second line of Frechet derivative matrix are different from zero )
$$
{f^-_{0.1}\over (f^-_{1.1})^2}[{1\over 4}D(T'_3)^t_{l1}+{1\over 4}f^-_{0.1}(T'_3)^t_{l2}-{1\over 4}f^-_{1.0}(T'_3)^t_{l3}]-{1\over f^-_{1.1}}[{1\over 4}f^+_{1.0}(T'_3)^t_{l1}+D(T'_3)^t_{l3}-f^-_{1.1}(T'_3)^t_{l5}]
$$
Result of simple algebraical calculation lead to explicit form of second row
\begin{equation}
\pmatrix{ 0 & 0 & \rig{f^+_{1.1}}{T_3} & -{3\over 4}{(f^-_{0.1})^2\over f^-_{1.1}}& D+{3\over 4}{f^-_{0.1}f^-_{1.0}\over f^-_{1.1}}  & {1\over 4}f^-_{0.1}D+{1\over 4}f^-_{1.1}f^+_{1.0}+{1\over 2}[(f^-_{0.1})'+{(f^-_{0.1})^2f^-_{1.0}\over f^-_{1.1}} \cr} \label{f2}
\end{equation}

After summation (\ref{f1}) and (\ref{f1}) with corresponding lines of (\ref{FII}) we obtain exactly first two lines of left hand side of (\ref{HIII}).

By similar calculation it is possible to 
check that (\ref{HIII}) is satisfied. Finally expression for $J^n_2$
\begin{equation}
\pmatrix{{1\over 2}f^+_{11}{1\over D}f^+_{11} & {1\over 4}f^+_{11}{1\over D}f^+_{10} & {1\over 4}f^+_{11}{1\over D}f^+_{01} & -{1\over 4}f^+_{11}{1\over D}f^-_{01} & -{1\over 4}f^+_{11}{1\over D}f^-_{10} & -{1\over 2}f^+_{11}{1\over D}f^-_{11} \cr   
{1\over 4}f^+_{10}{1\over D}f^+_{11} & {1\over 2}f^+_{10}{1\over D}f^+_{10} & -{1\over 4}f^+_{10}{1\over D}f^+_{01} & {1\over 4}f^+_{10}{1\over D}f^-_{01} & -{1\over 2}f^+_{10}{1\over D}f^-_{10} & -{1\over 4}f^+_{10}{1\over D}f^-_{11} \cr 
{1\over 4}f^+_{01}{1\over D}f^+_{11} & -{1\over 4}f^+_{01}{1\over D}f^+_{10} & {1\over 2}f^+_{01}{1\over D}f^+_{01} & -{1\over 2}f^+_{01}{1\over D}f^-_{01} & {1\over 4}f^+_{01}{1\over D}f^-_{10} & -{1\over 4}f^+_{01}{1\over D}f^-_{11} \cr 
-{1\over 4}f^-_{01}{1\over D}f^+_{11} & {1\over 4}f^-_{01}{1\over D}f^+_{10} & -{1\over 2}f^-_{01}{1\over D}f^+_{01} & {1\over 2}f^-_{01}{1\over D}f^-_{01} & -{1\over 4}f^-_{01}{1\over D}f^-_{10} & {1\over 4}f^-_{01}{1\over D}f^-_{11} \cr 
-{1\over 4}f^-_{10}{1\over D}f^+_{11} & -{1\over 2}f^-_{10}{1\over D}f^+_{10} & {1\over 4}f^-_{10}{1\over D}f^+_{01} & -{1\over 4}f^-_{10}{1\over D}f^-_{01} & -{1\over 2}f^-_{10}{1\over D}f^-_{10} & {1\over 4}f^-_{10}{1\over D}f^-_{11} \cr 
-{1\over 2}f^-_{11}{1\over D}f^+_{11} & -{1\over 4}f^-_{11}{1\over D}f^+_{10} & -{1\over 4}f^-_{11}{1\over D}f^+_{01} & {1\over 4}f^-_{11}{1\over D}f^-_{01} & {1\over 4}f^-_{11}{1\over D}f^-_{10} & {1\over 2}f^-_{11}{1\over D}f^-_{11} \cr}                      \label{J22}
\end{equation}
{\ref{J22}) and (\ref{J2}) solve the problem of the construction of the second Poisson structure
for the equations of 3-th waves hierarchy.
 
\section{Comments about multi-soliton solutions of the systems of 3-th waves hierarchy}

Let us find solution of the system equations of the second order (subsection 3 of section 2)
under additional condition $f^+_{i.j}=0$. The system under consideration looks as
$$
-\dot {f^-_{0.1}}=\nu_{0.1}(f^-_{0.1})'',\quad -\dot {f^-_{1.0}}=\nu_{1.0}(f^-_{1.0})'',
$$
$$
-\dot {f^-_{1.1}}=\nu_{1.1}(f^-_{1.1})''+\gamma_{1.1}f^-_{1.0}(f^-_{0.1})'+\delta_{1.1}f^-_{0.1}(f^-_{1.0})').
$$
where $\nu_{11}={\nu_{10}+\nu_{01}\over 4},\delta_{1.1}={3\nu_{10}-\nu_{01}\over 8},\gamma_{1.1}={\nu_{10}-3\nu_{01}\over 8}$.

Solution of two first linear equation are obvious
$$
f^-_{0.1}=\int d\mu e^{-\nu_{01} t \mu^2+\mu x} q(\mu),\quad f^-_{1.0}=\int d\lambda e^{-\nu_{10} t \lambda^2+\lambda x} p(\lambda)
$$
Let us find partial solution on nonhomogineous third equation in a form
$$  
f^-_{1.1}=\int d\mu \int d\lambda e^{-(\nu_{01} \mu^2+\nu_{10}\lambda^2)t+(\mu+\lambda) x} r(\mu,\lambda)
$$
After substitution this anzats for $f^-_{1.1}$  and obtained above solutions for $f^-_{1.0},f^-_{0.1}$ into third equation we come to an equality
$$
[\nu_{01} \mu^2+\nu_{10}\lambda^2-{\nu_{10}+\nu_{01}\over 4}(\mu+\lambda)^2] r(\mu,\lambda)=
(\gamma_{1.1}\mu+\delta_{1.1}\lambda)q(\mu)p(\lambda)
$$
Quadratical multiplier of the left side is equal to $2(\lambda-\mu)(\gamma_{1.1}\mu+\delta_{1.1}\lambda)$ and finally we obtain
$$
r(\mu,\lambda)={1\over 2}{q(\mu)p(\lambda)\over \lambda-\mu}
$$
and
$$  
f^-_{1.1}=\int d\mu \int d\lambda e^{-(\nu_{01} \mu^2+\nu_{10}\lambda^2)t+(\mu+\lambda) x} 
{1\over 2}{q(\mu)p(\lambda)\over \lambda-\mu}
$$
The last expression coincides (up to nonessential multiplier ${1\over 2}$) with solution in the case of 3-th problem. 
But in resolving of equations of discrete transformation only differentiation with respect to space coordinate take place. Thus all equations of discrete transformation in the case of 3-th wave problem and in the case under consideration will be have the same solution except of time dependent multiplier. And the form of multi soliton solution (up to this factor) will be the same for all systems of 3-th waves hierarchy.

\section{Outlook}

The main result of the present paper the explicit expression for second Poisson structure
(\ref{J2}) and (\ref{J22}), which allow to construct Hamiltonian reccurent operator and obtain all equations of 3-th wave hierarchy in explicit form. From the physical point of view these systems may be considered as three interacting fields of nonlinear Schredinger hierarchy connected with $A_1$ algebra.
All equations depended on one arbitrary numerical parameter, which can be connected with 
parameters of the particles of the fields describing by nonlinear Schredinger hierarchy.

The discrete transformation for n-wave problem in the case of arbitrary semisimple algebra
was presented in \cite{J} form and author have no doubts that the problem of equations of n-wave hierarchy and its multi-soliton solution may be resolved in explicit form. 

The most riddle to the author remain question about the nature of the group of discrete transformation. As it follows from its introduction \cite{J} it has some connection with the Weil group of the root space of semisimple algebra. Weil group is discrete one but non commutative. Discrete transformation in the case of the present paper is some reduction from group of discrete transformation of four dimensional self-dual Yang-Mills equations \cite{YM}. And thus understanding this situation in the case n-waves interaction will give possible guess to solution of this problem in Yang-Mills case.

\section*{Aknowledgements}

The author thanks CONACYT for financial support.  

\section{Appendix}

We present here different from zero matrix elements of Frechet derivative for $T_3$ discrete transformation -  $6\times 6$ matrix operator (see Introduction). All calculations are
done in connection with its definition (\ref{FR}) and explicit formulae for $T_3$ discrete transformation presented below.

\subsection{Discrete transformation $T_3$}
$$
\rig{f^+_{1.1}}{T_3}={1\over f^-_{1.1}},\quad \rig{f^+_{1.0}}{T_3}=-{f^-_{0.1}\over f^-_{1.1}},\quad \rig{f^+_{0.1}}{T_3}={f^-_{1.0}\over f^-_{1.1}},
$$
$$
\rig{f^-_{0.1}}{T_3}=-2(f^-_{0.1})'-f^-_{1.1}f^+_{1.0}-{1\over 2}{(f^-_{0.1})^2f^-_{1.0}\over f^-_{1.1}}+{(f^-_{1.1})'f^-_{0.1}\over f^-_{1.1}}
$$
$$
\rig{f^-_{1.0}}{T_3}=-2(f^-_{1.0})'+f^-_{1.1}f^+_{0.1}+{1\over 2}{(f^-_{1.0})^2f^-_{0.1}\over f^-_{1.1}}+{(f^-_{1.1})'f^-_{1.0}\over f^-_{1.1}}
$$
$$
\rig{f^-_{1.1}f^+_{1.1}}{T_3}=f^+_{1.1}f^-_{1.1}+(\ln f^-_{1.1})''+{1\over 2}{(f^-_{0.1})'f^-_{0.1}-(f^-_{1.0})'f^-_{0.1})\over f^-_{1.1}}+{1\over 2}(f^+_{1.0}f^-_{1.0}+f^+_{0.1}f^-_{0.1})+{1\over 4}({f^-_{1.0}f^-_{0.1}\over f^-_{1.1}})^2
$$
\subsection{Frechet derivative $\rig{F'}{T_3}$}

We present below different from zero matrix elements of Frechet operator
$$
\rig{F'_{16}}{T_3}=-{1\over (f^-_{1.1})^2},\quad \rig{F'_{24}}{T_3}=-{1\over f^-_{1.1}},\quad \rig{F'_{26}}{T_3}={f^-_{0.1}\over (f^-_{1.1})^2},\quad
\rig{F'_{35}}{T_3}={1\over f^-_{1.1}},\quad \rig{F'_{36}}{T_3}=-{f^-_{1.0}\over (f^-_{1.1})^2},
$$
$$
\rig{F'_{42}}{T_3}=-f^-_{1.1},\quad \rig{F'_{44}}{T_3}=-2D-Z+{(f^-_{1.1})'\over f^-_{1.1}},
\quad \rig{F'_{45}}{T_3}=-{(f^-_{0.1})^2\over 2f^-_{1.1}},
$$
$$
\rig{F'_{46}}{T_3}=-f^+_{1.0}+{f^-_{0.1}\over 2f^-_{1.1}}Z+{f^-_{0.1}\over f^-_{1.1}}D-{f^-_{0.1}(f^-_{1.1})'\over (f^-_{1.1})^2},
$$
$$
\rig{F'_{53}}{T_3}=f^-_{1.1},\quad \rig{F'_{54}}{T_3}={(f^-_{1.0})^2\over 2f^-_{1.1}},\quad \rig{F'_{55}}{T_3}=-2D+Z+{(f^-_{1.1})'\over f^-_{1.1}},
$$
$$
\rig{F'_{56}}{T_3}=f^+_{0.1}-{f^-_{1.0}\over 2f^-_{1.1}}Z+{f^-_{1.0}\over f^-_{1.1}}D-{f^-_{1.0}(f^-_{1.1})'\over (f^-_{1.1})^2},
$$
$$
\rig{F'_{61}}{T_3}=(f^-_{1.1})^2,\quad \rig{F'_{62}}{T_3}={1\over 2}f^-_{1.0}f^-_{1.1},\quad \rig{F'_{63}}{T_3}={1\over 2}f^-_{0.1}f^-_{1.1},\quad \rig{F'_{64}}{T_3}={1\over 2}(f^-_{1.0}D-(f^-_{1.0})')+{1\over 2}f^-_{1.0}Z+{1\over 2}f^-_{1.1}f^+_{0.1},
$$
$$
\rig{F'_{65}}{T_3}=-{1\over 2}(f^-_{0.1}D-(f^-_{0.1})')+{1\over 2}f^-_{0.1}Z+{1\over 2}f^-_{1.1}f^+_{1.0}
$$
$$
\rig{F'_{66}}{T_3}=D^2-2{(f^-_{1.1})'\over f^-_{1.1}}D+{(f^-_{1.1})'(f^-_{1.1})'\over f^-_{1.1}}+2f^-_{1.1}f^+_{1.1}+{1\over 2}(f^-_{1.0}f^+_{1.0}+f^-_{0.1}f^+_{0.1})-{1\over 4}Z^2
$$
where $Z={f^-_{1.0}f^-_{0.1}\over f^-_{1.1}}$.

\end{document}